\documentclass[%
 reprint,
 superscriptaddress,
nofootinbib,
 amsmath,amssymb,
 aps,
 prx,
]{revtex4-2}

\usepackage{graphicx}
\usepackage{dcolumn}
\usepackage{bm}
\usepackage{hyperref}
\hypersetup{colorlinks=true,linkcolor=blue,citecolor=blue,filecolor=blue,urlcolor=blue}
\usepackage{multirow}
\usepackage{makecell}
\usepackage{booktabs}
\usepackage{physics}

\usepackage{bbold}

\begin{document}

\title{
Data-Efficient Quantum Noise Modeling via Machine Learning
}

\author{Yanjun Ji}%
 \email{y.ji@fz-juelich.de}
\affiliation{%
Institute for Quantum Computing Analytics (PGI-12), Forschungszentrum Jülich, 52425 Jülich, Germany
}%

\author{Marco Roth}%
\email{marco.roth@ipa.fraunhofer.de}
\affiliation{%
Fraunhofer Institute for Manufacturing Engineering and Automation (IPA), Nobelstrasse 12, 70569 Stuttgart, Germany
}%

\author{David A. Kreplin}%
\affiliation{%
Heilbronn University of Applied Sciences, Max-Planck-Str. 39, 74081 Heilbronn, Germany
}%

\author{Ilia Polian}
\affiliation{Institute of Computer Architecture and Computer Engineering, University of Stuttgart, Pfaffenwaldring 47, 70569 Stuttgart, Germany}

\author{Frank K. Wilhelm}
\affiliation{%
Institute for Quantum Computing Analytics (PGI-12), Forschungszentrum Jülich, 52425 Jülich, Germany
}%
\affiliation{%
Theoretical Physics, Saarland University, 66123 Saarbrücken, Germany
}%

\date{\today}

\begin{abstract}

Maximizing the computational utility of near-term quantum processors requires predictive noise models that inform robust, noise-aware compilation and error mitigation.
Conventional models often fail to capture the complex error dynamics of real hardware or require prohibitive characterization overhead. We introduce a data-efficient framework that first constructs a physically motivated, parameterized noise model, and subsequently employs machine learning-driven Bayesian optimization to identify its parameters. Our approach circumvents costly characterization protocols by estimating algorithm- and hardware-specific error parameters directly from readily available experimental data derived from existing application and benchmark circuit executions.
The generality and robustness of the framework are demonstrated across diverse algorithms and superconducting devices, yielding high-fidelity predictions by estimating an independent parameter set tailored to each specific algorithm-hardware context. Crucially, we show that a model calibrated exclusively on small-scale circuits accurately predicts the behavior of larger validation circuits. Our data-efficient approach achieves up to a 65\% improvement in model fidelity quantified by the Hellinger distance between predicted and experimental circuit output distributions, compared to standard noise models derived from device properties.
This work establishes a practical paradigm for application-aware noise characterization, enabling compilation and error-mitigation strategies tailored to the specific interplay between quantum algorithms and device-specific noise dynamics.

\end{abstract}

\maketitle

\section{Introduction}

Quantum computing offers a significant opportunity to address complex optimization problems. In the current noisy intermediate-scale quantum era~\cite{preskill2018quantum}, variational algorithms exemplified by the quantum approximate optimization algorithm (QAOA)~\cite{farhi2014quantum} and the variational quantum eigensolver (VQE)~\cite{peruzzo2014variational} have been explored for diverse applications, including portfolio optimization~\cite{brandhofer2023benchmarking}, unit commitment in power systems~\cite{koretsky2021adapting}, and quantum chemistry~\cite{kandala2017hardware}.
Superconducting circuits~\cite{clarke2008superconducting, krantz2019quantum} have emerged as a leading platform because they benefit from established microchip manufacturing techniques and offer fast operating times.
However, the practical utility of these devices is constrained by noise mechanisms and hardware limitations that hinder algorithm's performance.
To enhance the reliability of quantum applications on such devices, noise-aware compilation has become an essential strategy.
Quantum compilation translates quantum algorithms into sequences of physical operations that can be executed on a target backend. This process ensures that algorithms satisfy the hardware's intrinsic constraints and optimizes the circuit to minimize the impact of errors. Specifically, the compilation needs to adapt the algorithm to the device's limited qubit connectivity and small set of supported gates, also referred to as basis gates, by inserting SWAP gates and decomposing operations in algorithms into the device's basis gate set. This is enabled by compilers supported by, e.g., Qiskit \cite{qiskit}, TKET \cite{sivarajah2020tket}, and Munich quantum toolkit \cite{burgholzer2025munich, burgholzer2025MQTCore}.
Beyond ensuring basic executability, noise-aware compilation aims to maximize the fidelity of the compiled circuit. To achieve this, the compiler needs an accurate model of the device's error characteristics, which often vary over time \cite{ji2022calibration}. This model guides strategic decisions, including selecting a subset of high-fidelity qubits for execution, optimizing the routing of logical qubits to minimize noise accumulation \cite{quetschlich2025mqt, ji2025algorithm, montanez2025optimizing}, and choosing between different gate sequences that, while logically equivalent, may exhibit varying resilience to noise \cite{ji2023improving} and other imperfections like Trotter errors \cite{faehrmann2022randomizing, childs2021theory, childs2019faster}.

An accurate and predictive noise model is essential for achieving effective noise-aware compilation. However, constructing error models for real hardware presents a significant trade-off between model fidelity and experimental cost.
Traditional error models, such as those offered by hardware vendors, typically assume independent Pauli noise channels that fail to capture the intricate dynamics inherent to real hardware.
In practice, superconducting quantum devices exhibit spatially and temporally correlated errors \cite{maciejewski2021modeling, harper2020efficient, gambetta2012characterization}, non-Markovian memory effects \cite{agarwal2024modelling, zhang2022predicting, white2020demonstration}, and time-varying decoherence parameters \cite{etxezarreta2021time}. 
Conversely, while comprehensive characterization techniques like quantum process tomography \cite{poyatos1997complete} can capture these intricate effects, they are resource-intensive and impractical to scale for routine calibration of larger quantum processors.
Various intermediate strategies have been explored to bridge this gap. 
For instance, some approaches use optimization techniques such as genetic algorithms to build models for specific error types \cite{georgopoulos2021modeling}, while others propose scalable Markovian models that often require custom hardware tuning to be accurate \cite{brand2024markovian}.
Further studies have shown that noise models should be tailored to specific applications, as no single model captures all noise dynamics \cite{dahlhauser2021modeling}. Even integrated approaches that combine Markovian and non-Markovian effects to improve accuracy still tend to rely on resource-intensive characterizations \cite{oda2023noise}. These studies highlight the persistent trade-off between model fidelity and experimental overhead, emphasising the necessity for a more efficient approach.

\begin{figure}[tb]
    \centerline{\includegraphics[width=\columnwidth]{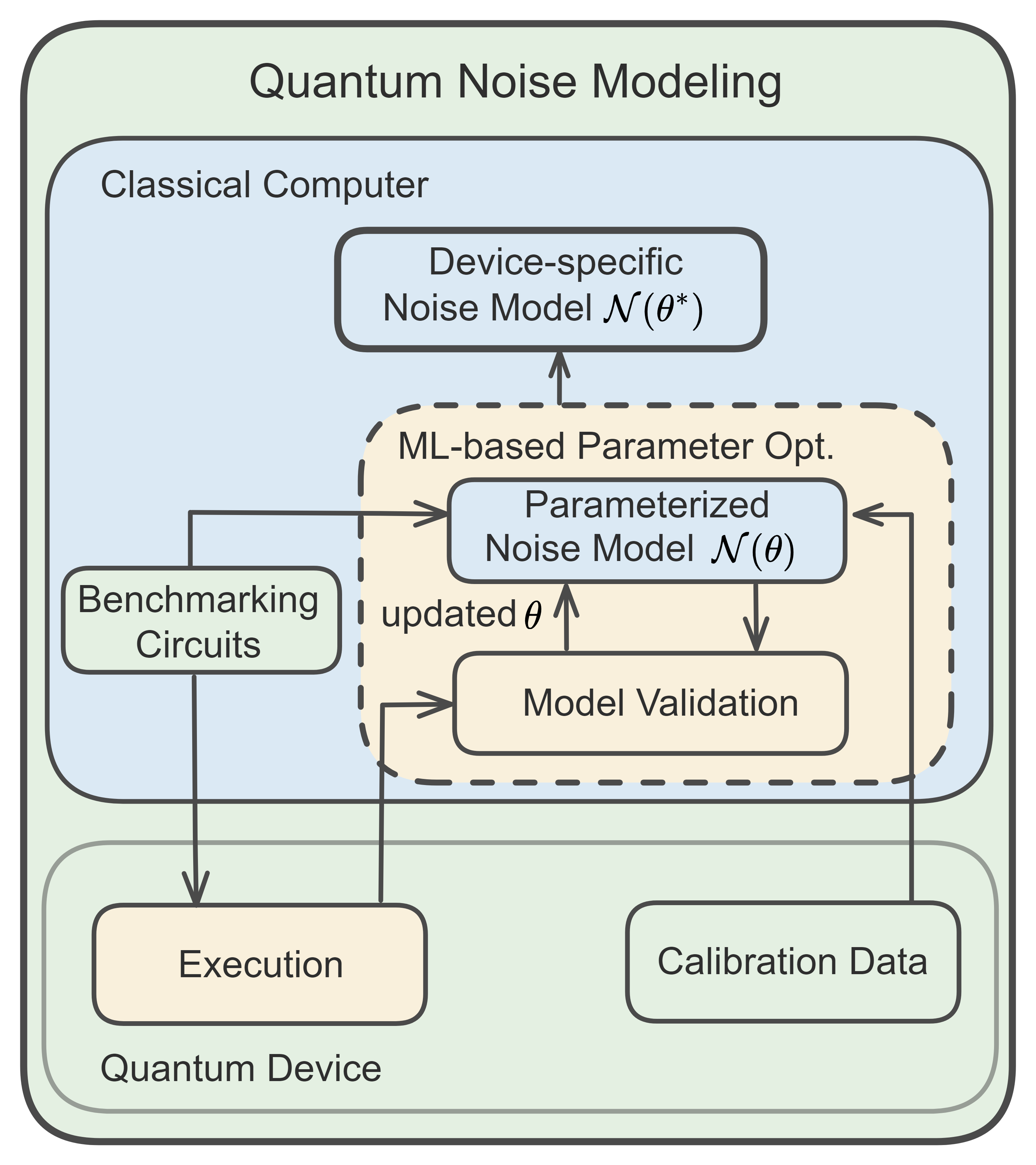}}
    \caption{Flowchart of the data-efficient quantum noise modeling framework. A parameterized noise model $\mathcal{N}(\boldsymbol{\theta})$ is initially constructed based on calibration data provided by hardware vendors.
    The device-specific noise model $\mathcal{N}(\boldsymbol{\theta^*})$ is then identified by executing benchmarking circuits on real quantum hardware and aligning experimental results with noisy simulations. This iterative optimization, performed on a classical computer, leverages machine learning (ML)-based Bayesian optimization to update $\boldsymbol{\theta}$. This approach is highly flexible, designed to ingest outcome distributions from diverse preexisting experimental data, including application-specific circuits, avoiding the need for dedicated characterization.}
    \label{fig:flowchat_noise_modeling}
\end{figure}

To address these limitations, we introduce a practical and data-efficient framework that circumvents resource-intensive characterization protocols by leveraging readily available experimental data, such as application benchmarks. Our core contribution is a circuit-size-independent parameterized noise model $\mathcal{N}(\boldsymbol{\theta})$, whose parameters are determined via iterative, ML-driven optimization, as depicted in Fig.~\ref{fig:flowchat_noise_modeling}. By minimizing the Hellinger distance~\cite{hellinger1909neue} between simulated and experimental output distributions, we estimate optimal parameters $\boldsymbol{\theta}^*$ directly from existing circuit executions. Crucially, this framework functions as an application-aware noise characterization. Rather than positing a single universal error profile, we estimate independent parameter sets tailored to specific algorithm-device contexts (e.g., QAOA, VQE, or random circuits on a target hardware). This approach captures the interplay between circuit structure and device-intrinsic error dynamics, yielding optimized models that significantly outperform default hardware characterizations across diverse IBM quantum processors.

We demonstrate scalability by training on 4--6 qubit benchmarks and accurately predicting out-of-distribution behavior in 7--9 qubit circuits. This approach achieves a substantial model accuracy improvement, reducing the mean Hellinger distance to experimental results by 50\% (65\% peak reduction). Such predictive accuracy on larger-scale systems advances beyond foundational studies restricted to small-scale demonstrations~\cite{georgopoulos2021modeling}, establishing that high-fidelity, device-specific noise models can be learned from compact datasets. This enables practitioners to continuously refine noise models using existing data, advancing noise-aware compilation and error mitigation~\cite{cai2023quantum, kandala2019error, ji2024synergistic} without additional experimental cost.

The paper is structured as follows. Section~\ref{sec:theory} presents the theoretical framework and methodology, establishing the baseline provided by device-calibrated noise models and introducing our parameterized noise model. It also details the machine learning (ML)-based Bayesian optimization (BO) technique employed for parameter determination and concludes with a critical analysis of the framework's scalability. Section~\ref{sec:results} demonstrates a thorough experimental validation of our approach, benchmarking the model's predictive accuracy against data from multiple superconducting processors using a diverse set of applications, including the QAOA \cite{farhi2014quantum}, VQE \cite{peruzzo2014variational}, and random circuits. This validation includes an in-depth single-device characterization, a cross-platform comparative analysis to demonstrate hardware generality, and a study of training efficiency. Finally, Sec.~\ref{sec:conclusion} concludes the paper.

\section{Theoretical Framework and Methodology\label{sec:theory}}

This section details our noise modeling methodology. We first introduce a baseline noise model for comparison.
We then introduce our circuit-size-independent parameterized noise model, followed by parameter determination strategy. Finally, we discuss the scalability of our approach.

\subsection{Default noise model}

Most providers of superconducting quantum processing units (QPUs) offer noise models, which are often derived directly from the quantum device's calibration data. These data, provided and periodically updated by the hardware vendors, quantify the intrinsic properties of the qubits, specifically their relaxation ($T_1$) and dephasing ($T_2$) times, as well as the error rates and execution durations of the device’s supported basis gates. A common example is the standard model available in Qiskit~\cite{qiskit} for IBM devices, which can be constructed using the provided backend configurations and properties.
This model isolates errors into two primary categories: gate errors and readout errors.
Gate errors are modeled as a composed quantum channel $\mathcal{E}_{\text{gate}}$ applied after each ideal unitary operation. This channel is formed by the composition of a depolarizing channel $\mathcal{E}_{\text{depol}}$, followed by a thermal relaxation channel $\mathcal{E}_{\text{thermal}}$:
\begin{equation}
    \mathcal{E}_{\text{gate}} = \mathcal{E}_{\text{thermal}} \circ \mathcal{E}_{\text{depol}},
\end{equation}
where the composition $\mathcal{A} \circ \mathcal{B}$ implies that channel $\mathcal{B}$ is applied first.
The thermal relaxation channel $\mathcal{E}_{\text{thermal}}$ models energy relaxation ($T_1$) and dephasing ($T_2$) processes occurring during the gate's finite duration. The depolarizing channel $\mathcal{E}_{\text{depol}}$ is not an independently measured quantity; rather, its error probability is chosen such that the infidelity of the composite channel $\mathcal{E}_{\text{gate}}$ precisely matches the experimentally characterized gate error rate $e_g$.

Readout errors in an $n$-qubit system are modeled via a $2^n \times 2^n$ stochastic matrix $M$, where $M_{ij}$ denotes the conditional probability $P(\text{measure } \ket{i} \mid \text{true state } \ket{j})$ of measuring computational basis state $\ket{i}$ given true state $\ket{j}$. These errors, treated as a classical post-processing step, transform the ideal probability distribution $P_{\text{ideal}}$ into the observed distribution $P_{\text{meas}} = M P_{\text{ideal}}$, with $M$ derived from empirical measurements.

Although this baseline model can be easily obtained, it excludes cross-talk and other correlated effects. Moreover, it captures only noise mechanisms explicitly characterized during routine device calibration procedures, creating a static model that cannot account for the subsequent temporal evolution of the physical device.
To better capture the real noise behavior of quantum algorithms executing on QPUs, we develop a tractable, data-efficient error model, which we detail below.

\subsection{Parameterized noise model}

Our model is designed to be parameterized by a vector of learnable coefficients $\boldsymbol{\theta}$, which are then chosen using ML-based optimization strategy to minimize the discrepancy between simulation and experimental data.
An error channel $\mathcal{E}_g$ for any gate operation is modeled as the composition of several distinct error channels, each targeting a specific physical error mechanism.

\subsubsection{Single-qubit gate errors}
For a single-qubit gate $G_q$ (e.g., $X$, $SX$) in the bases gate set with execution duration $t_g$ and error rate $e_g$, the error channel $\mathcal{E}_{\text{1Q}}$ is modeled as a composition of four channels:
\begin{equation}
\mathcal{E}_{\text{1Q}} = \mathcal{E}_{\text{ph}}  \circ \mathcal{E}_{\text{amp}} \circ \mathcal{E}_{\text{dep}} \circ \mathcal{E}_{\text{coh}},
\end{equation}
where coherent errors are applied first, followed sequentially by depolarizing, 
amplitude damping, and phase damping channels.
The coherent error channel $\mathcal{E}_{\text{coh}}$ models systematic over- or under-rotations due to control pulse distortions. This channel acts on the density matrix $\rho$ via $\mathcal{E}_{\text{coh}}(\rho) = U_{\text{coh}}~\rho~ U_{\text{coh}}^\dagger$, defined by the unitary operator $U_{\text{coh}} = \exp(-i H_{\text{coh}} t_g)$. The corresponding error Hamiltonian $H_{\text{coh}}$ is a linear combination of Pauli operators with learnable amplitudes $\theta_{x,y,z}$:
\begin{equation}
H_\mathrm{coh} = \theta_x X + \theta_y Y + \theta_z Z.
\end{equation}
For the identity gate, this coherent error is omitted.

The depolarizing error channel $\mathcal{E}_{\text{dep}}$ models a generic, incoherent disruption of the quantum state.
For an $n$-qubit system, the depolarizing channel is given by
\begin{equation}
  \mathcal{E}_{\text{dep}}^{(n)}(\rho)
  = (1 - \lambda_{\text{dep}}^{(n)})\,\rho
    + \lambda_{\text{dep}}^{(n)}\, \frac{I}{2^n},
\end{equation}
where $\lambda_{\text{dep}}^{(n)}$ is the depolarizing noise strength. For single-qubit, the noise strength $\lambda_{\text{dep}}$ is defined as a duration-aware, saturating function
\begin{equation}
    \lambda_{\text{dep}}(e_g, t_g) = 1 - e^{-(k_{\text{dep}} e_g + b_{\text{dep}}) t_g/t_{\text{char}}}.
    \label{eq:prob_dep_singl_q}
\end{equation}
Here, $k_{\text{dep}}$ and $b_{\text{dep}}$ are learnable parameters that scale and offset the error rate, and $t_{\text{char}}$ is a constant characteristic gate time for the device defined as the median of all single qubit gate durations. This form captures the physical reality that error probability saturates for long-duration operations.

The amplitude damping channel $\mathcal{E}_{\text{amp}}$ models energy relaxation. Its action on the density matrix $\rho$ is defined by the Kraus decomposition
\begin{equation}
    \mathcal{E}_{\text{amp}}(\rho) = \sum_{k=0}^{1} K_k^{\text{amp}} \rho (K_k^{\text{amp}})^\dagger,
\end{equation}
with Kraus operators
\begin{equation}
    K_0^{\text{amp}} = \begin{pmatrix} 1 & 0 \\ 0 & \sqrt{1 - \lambda_{\text{amp}}} \end{pmatrix}, \quad 
K_1^{\text{amp}} = \begin{pmatrix} 0 & \sqrt{\lambda_{\text{amp}}} \\ 0 & 0 \end{pmatrix}.
\label{eq:kraus_ope_amp}
\end{equation}
The relaxation probability $\lambda_{\text{amp}}$ is governed by $T_1$ and $t_g$, with an added learnable offset $b_{\text{amp}}$:
\begin{equation}
    \lambda_{\text{amp}}(t_g, T_1) = 1 - e^{-t_g/T_1} + b_{\text{amp}}.
    \label{eq:ampli_dampi_chann}
\end{equation}

To capture non-Markovian dephasing, potentially arising from low-frequency noise sources like two-level-system fluctuators, we model phase damping channel $\mathcal{E}_{\text{ph}}$ using a stretched exponential decay.
Similarly, the corresponding Kraus operators are 
\begin{equation}
    K_0^{\text{ph}} =  \begin{pmatrix} 1 & 0 \\ 0 & \sqrt{1-\lambda_{\phi}} \end{pmatrix}, \quad 
K_1^{\text{ph}} = \begin{pmatrix} 0 & 0 \\ 0 & \sqrt{\lambda_{\phi}} \end{pmatrix}.
\label{eq:kraus_ope_phase}
\end{equation}
The dephasing parameter $\lambda_{\phi}$ is given by
\begin{equation}
    \lambda_{\phi}(t_g, T_\phi) = 1 - e^{-2 \left(t_g/T_\phi\right)^{\beta_{\text{1Q}}}},
    \label{eq:prob_ph_singl_q}
\end{equation}
where $T_\phi$ is the pure dephasing time derived from the measured $T_1$ and $T_2$, satisfying $1/T_2 = 1/(2T_1) + 1/T_\phi$, and $\beta_{\text{1Q}}$ is a learnable exponent.

\subsubsection{Two-qubit gate errors}

For a two-qubit basis gate $G_{2Q}$ acting on the qubit pair $(q_0, q_1)$, with gate duration $t_{2Q}$ and error rate $e_{2Q}$, the corresponding error channel $\mathcal{E}_{\text{2Q}}$ follows a similar compositional structure but includes terms for capturing crosstalk and correlated errors:
\begin{equation}
\mathcal{E}_{\text{2Q}} = \mathcal{E}_{{zz}} \circ (\mathcal{E}_{\text{ph}} \otimes \mathcal{E}_{\text{ph}}) \circ (\mathcal{E}_{\text{amp}} \otimes \mathcal{E}_{\text{amp}}) \circ \mathcal{E}_{\text{dep}}^{(2)} \circ \mathcal{E}_{\text{coh}}^{(2)},
\end{equation}
composing the two-qubit coherent, two-qubit depolarizing, local amplitude and phase damping, and correlated ZZ dephasing error channels.
Two-qubit coherent error $\mathcal{E}_{\text{coh}}^{(2)}$ models coherent crosstalk, such as residual $ZZ$ coupling or control crosstalk. This error channel acts as $\mathcal{E}_{\text{coh}}^{(2)}(\rho) = U_{\text{coh}}^{(2)} \rho (U_{\text{coh}}^{(2)})^\dagger$ with $U_{\text{coh}}^{(2)} = \exp(-i H_{\text{coh}}^{(2)} t_{2Q})$. The error Hamiltonian includes two-qubit Pauli terms:
\begin{equation}
    H_{\text{coh}}^{(2)} = \theta_{ix} I \otimes X + \theta_{zx} Z \otimes X + \theta_{zz} Z \otimes Z,
    \end{equation}
where $\theta_{ix,zx,zz}$ are learnable parameters.

Analogous to the single-qubit case in Eq.~\eqref{eq:prob_dep_singl_q}, the two-qubit depolarizing channel $\mathcal{E}_{\text{dep}}^{(2)}$ is characterized by a duration-aware, saturating probability $\lambda_{\text{dep}}^{(2)}$, given by
\begin{equation}
    \lambda_{\text{dep}}^{(2)}(e_{2Q}, t_{2Q}) = 1 - e^{-(k_{\text{dep,2Q}} e_g + b_{\text{dep,2Q}}) t_{2Q}/t_{\text{char,2Q}}},
\end{equation}
where $k_{\text{dep,2Q}}$ and $b_{\text{dep,2Q}}$ are learnable parameters. The characteristic timescale $t_{\text{char,2Q}}$ is fixed as the median duration of the two-qubit gates.

The single-qubit amplitude damping and phase dephasing channels $\mathcal{E}_{\text{amp}}$ and $\mathcal{E}_{\text{ph}}$ are applied independently to the qubit pair ($q_0$, $q_1$), using Kraus operators defined in Eqs.~\eqref{eq:kraus_ope_amp} and \eqref{eq:kraus_ope_phase}, i.e., 
\begin{equation}
    \mathcal{E}_{\text{amp}} \otimes \mathcal{E}_{\text{amp}}(\rho) = \sum_{k,l=0}^{1} (K_k^{\text{amp}} \otimes K_l^{\text{amp}}) \rho (K_k^{\text{amp}} \otimes K_l^{\text{amp}})^\dagger,
\end{equation}
and similarly for $\mathcal{E}_{\text{ph}} \otimes \mathcal{E}_{\text{ph}}$.
Their respective error strengths $\lambda_{\text{amp},2Q}$ and $\lambda_{\phi,2Q}$ are defined as
\begin{align}
    \lambda_{\text{amp},2Q}(t_{2Q}, T_1) &= 1 - e^{-t_{2Q}/T_1} + b_{\text{amp},2Q},\\
    \lambda_{\phi,2Q}(t_{2Q}, T_\phi) &= 1 - e^{-2(t_{2Q}/T_\phi)^{\beta_{2Q}}} + b_{\phi,2Q}.
\end{align} These interactions are quantified by the learnable parameters $b_{\text{amp,2Q}}$, $\beta_{\text{2Q}}$, and $b_{\phi\text{,2Q}}$. Here, the offset parameter $b_{\phi\text{,2Q}}$ is additionally introduced for two-qubit gates, compared to single qubit gates (Eq.~\eqref{eq:prob_ph_singl_q}), to better characterize the more complex error mechanisms of two-qubit gates.
We note that $\lambda_{\text{amp},2Q}$ and $\lambda_{\phi,2Q}$ are evaluated independently for $q_0$ and $q_1$ based on their respective coherence times ($T_1, T_\phi$), yielding asymmetric local noise channels despite the common gate duration $t_{2Q}$.

The correlated $ZZ$ dephasing channel $\mathcal{E}_{{zz}}$ introduces a correlated dephasing error, a common mechanism in superconducting qubits arising from spectator qubit interactions or flux noise. It is modeled as a Pauli channel that applies a $Z \otimes Z$ operator with probability $\lambda_{zz}$:
\begin{equation}
    \mathcal{E}_{{zz}}(\rho) = (1 - \lambda_{zz}) \rho + \lambda_{zz} (Z \otimes Z) \rho (Z \otimes Z).
    \end{equation}
The parameter $\lambda_{zz}$ is also modeled as a saturating, duration-dependent function parameterized by a learnable rate $k_{zz}$:
\begin{equation}
    \lambda_{zz}(t_{2Q}) = 1 - e^{-k_{zz} t_{2Q}/t_{\text{char,2Q}}}.
\end{equation}

\subsubsection{Readout errors}

We extend the standard model of independent single-qubit readout errors to include a scalable approximation of correlated errors. In superconducting processors, readout correlations often arise from shared resonators or multiplexed electronics, potentially linking multiple qubits. However, modeling the full joint readout distribution of $n$ qubits requires a confusion matrix scaling as $2^n \times 2^n$. This exponential scaling is incompatible with our objective of a data-efficient, circuit-size-independent framework. Instead, we model correlations at the pairwise level as the leading-order correction to the independent model. We approximate the global readout confusion matrix $\mathcal{M}_{\text{global}}$ as a tensor product of independent single-qubit matrices $M^{(1)}$ and disjoint two-qubit matrices $M^{(2)}$:
\begin{equation}
    \mathcal{M}_{\text{global}} \approx \left( \bigotimes_{(q_i, q_j) \in \mathcal{S}} M_{ij}^{(2)} \right) \otimes \left( \bigotimes_{q_k \notin \mathcal{S}} M_{k}^{(1)} \right),
    \label{eq:readout_tensor}
\end{equation}
where $\mathcal{S}$ is a set of disjoint qubit pairs. To construct $\mathcal{S}$, we employ a deterministic greedy maximal matching algorithm on the device coupling graph. This algorithm iterates through the edges of the coupling map (sorted by qubit index to ensure reproducibility) and selects a pair $(q_i, q_j)$ if neither qubit has been previously selected. This approach maximizes the coverage of correlated errors, prioritizing physically connected neighbors where crosstalk is typically strongest, while ensuring each qubit participates in at most one correlated channel.

For a selected pair $(q_0, q_1)$, we begin with the $4 \times 4$ confusion matrix representing independent errors, $M_{\text{ind}} = M_{q_1} \otimes M_{q_0}$. We then construct the correlated confusion matrix $M_{ij}^{(2)}$ by injecting correlations into this product. Specifically, we model the $|00\rangle \leftrightarrow |11\rangle$ correlation, corresponding to a simultaneous correlated readout misassignment on both qubits. This is achieved by transferring a probability mass $\lambda_{\text{corr}}^{00 \leftrightarrow 11}$ from the diagonal elements to the anti-diagonal elements (correlated error) within the subspaces of the affected states, such as
\begin{equation}
    M_{0,0} \rightarrow M_{0,0} - \lambda_{\text{corr}} \quad \text{and} \quad
    M_{0,3} \rightarrow M_{0,3} + \lambda_{\text{corr}}.
\end{equation}
Similarly, the $|01\rangle \leftrightarrow |10\rangle$ correlation models state-swapping errors, where a probability $\lambda_{\text{corr}}^{01 \leftrightarrow 10}$ is transferred between elements corresponding to $\ket{01}$ and $\ket{10}$ states. The magnitude of the correlation is scaled by the average local error rate of the pair, $\bar{e} = (e_{q_0} + e_{q_1})/2$. The probability of a correlated error is parameterized as a linear function:
\begin{equation}
\lambda_{\text{corr}} = a~\bar{e} + b,
\end{equation}
where separate parameters $(a, b)$ are learnable variables for $\ket{00} \leftrightarrow \ket{11}$ and $\ket{01} \leftrightarrow \ket{10}$ correlation types, respectively.

\begin{table}[tb]
\centering
\caption{The 20 learnable parameters $\boldsymbol{\theta}$ of the noise model, grouped by function role.}
\label{tab:noise_parameters}
\begin{tabular}{@{}lll@{}}
\toprule
{Parameter} & {Description} & {Type} \\
\midrule
\multicolumn{3}{c}{{\textbf{Single-Qubit Gate Parameters (7)}}} \\
$k_{\text{dep}}$    & Depolarizing scaling factor         & Stochastic         \\
$b_{\text{dep}}$    & Depolarizing offset        & Stochastic         \\
$b_{\text{amp}}$    & Amplitude damping offset        & Stochastic         \\
$\theta_{x,y,z}$          & Coherent rotation angles       & Coherent           \\
$\beta_{\text{1Q}}$    & Stretched dephasing exponent   & Non-Markovian      \\
\midrule
\multicolumn{3}{c}{{\textbf{Two-Qubit Gate Parameters (9)}}} \\
$k_{\text{dep,2Q}}$    & Depolarizing scaling factor         & Stochastic         \\
$b_{\text{dep,2Q}}$    & Depolarizing offset        & Stochastic         \\
$b_{\text{amp,2Q}}$    & Amplitude damping offset        & Stochastic         \\
$b_{\phi\text{,2Q}}$    & Phase damping offset       & Stochastic         \\
$\theta_{ix,zx,zz}$       & Coherent crosstalk angles         & Coherent           \\
$\beta_{\text{2Q}}$    & Stretched dephasing exponent   & Non-Markovian      \\
$k_{zz}$              & Correlated $ZZ$ dephasing rate & Correlated \\
\midrule
\multicolumn{3}{c}{{\textbf{Correlated Readout Parameters (4)}}} \\
$a,b_{00\leftrightarrow11}$ & Coefficients ($|00\rangle\leftrightarrow|11\rangle$) & Correlated \\
$a,b_{01\leftrightarrow10}$ & Coefficients ($|01\rangle\leftrightarrow|10\rangle$) & Correlated \\
\bottomrule
\end{tabular}
\end{table}

\subsubsection{Model parameterization and calibration integration}

The complete noise model, encompassing gate and readout error channels detailed above, is fully specified by a set of 20 free parameters $\boldsymbol{\theta}$. These parameters quantify the strength of each physical error mechanism, from coherent rotations and stochastic decoherence to non-Markovian dynamics and correlated readout, forming the target vector for our data-efficient optimization. For clarity, Table~\ref{tab:noise_parameters} summarizes these learnable model parameters, grouped by their functional role: 7 for single-qubit, 9 for two-qubit, and 4 for readout error characterization.

Crucially, our framework employs a hybrid parameterization strategy that integrates global learnable parameters with local calibration data.
Instead of learning independent error rates for every qubit, which would result in an unscalable parameter space, the 20 optimization variables $\boldsymbol{\theta}$ function as global modifiers applied to the spatially varying, qubit-specific calibration inputs. The error channels leverage standard device calibration parameters, including relaxation times $T_1$, dephasing times $T_2$, gate durations and error rates of basis gates, which are sourced directly from routine vendor-provided characterization data. Utilizing these existing data requires no dedicated calibration circuits or additional QPU time from the user.
This architecture ensures that the model naturally captures spatial heterogeneity. For instance, in the amplitude damping channel (Eq.~\ref{eq:ampli_dampi_chann}), the relaxation time $T_1$ varies locally per qubit, while the offset $b_{\text{amp}}$ is shared globally. Consequently, for a fixed gate duration $t_g$, a qubit with lower $T_1$ is assigned a higher damping probability than one with higher $T_1$ despite identical $b_{\text{amp}}$. Learning global corrections to local physical priors avoids the prohibitive complexity of learning unique parameter sets for every qubit and qubit pair, while successfully modeling the non-uniform error landscape.
Our approach therefore complements, rather than replaces, standard device characterization routines. It utilizes existing calibration data as a physically grounded baseline and employs ML to infer correction factors that refine the effective noise model using circuit-execution statistics captured within the specific algorithm–hardware operating regime.

\subsection{Parameter optimization}

The error model introduced in the previous section contains 20 free parameters $\boldsymbol{\theta}$.
Our goal is to choose the parameters of the extended model such that the model accurately reproduces sample distributions $P_i$ obtained from executing circuits $C_i\in\mathcal{C}$ on a specific quantum computing backend. This can be formalized as identifying the parameter configuration $\boldsymbol{\theta}^*$ that minimizes the Hellinger distance~\cite{hellinger1909neue} between simulation results and the results obtained on a real device.
The Hellinger distance $D_H$ is a bounded, symmetric metric for comparing discrete probability distributions and is numerically stable for finite-shot, sparse bitstring histograms. It ranges from 0 to 1, with 0 indicating identical distributions, and is defined as
\begin{equation}
    D_H(P, Q) = \left(1-\sum_j \sqrt{p_j q_j}\right)^{1/2},
    \label{eq:hellinger_distance}
\end{equation}
where $p_j$ and $q_j$ are probabilities of outcome $j$ in distributions $P$ and $Q$, respectively.
Specifically, we minimize the mean Hellinger distance across all outcome distributions produced by the circuits in the training set, with respect to the circuits of interest, to obtain
\begin{equation}
\boldsymbol{\theta}^*=\text{argmin}_{\boldsymbol{\theta}}\left\lbrace\frac{1}{n_{\rm tr}}\sum_{i=1}^{n_{\rm tr}} D_H[P_i, Q_i(\boldsymbol{\theta})]\right\rbrace\,.
\label{eq:theta_opt}
\end{equation}
Here, ${n_{\rm tr}}= \vert {\mathcal{C}}\vert$ is the size of the training set, $Q_i(\boldsymbol{\theta})$ is the empirical distribution obtained by simulating the circuit with a noise model with parameters $\boldsymbol{\theta}$.
Each distribution is estimated from finite-shot measurement outcomes. Throughout, we use $30.000$ shots (samples) per circuit in both the training and validation sets.
We emphasize that Eq.~\eqref{eq:theta_opt} defines a global fit, meaning that a single parameter vector $\boldsymbol{\theta}$ is identified jointly over all circuits in $\mathcal{C}$, with no circuit-specific parameters. This structure enforces that $\boldsymbol{\theta}$ captures systematic, context-consistent error mechanisms shared by the algorithm family, rather than memorizing individual circuit outcomes, which potentially mitigates overfitting.

Since the configuration space of the model is large and evaluating the Hellinger distance requires running noisy simulations, identifying $\boldsymbol{\theta}^*$ is not tractable using brute force search.
Moreover, gradient-based optimization faces practical limitations in this setting. Efficient gradient-based optimization via automatic differentiation would require end-to-end differentiability of the noisy-simulation stack with respect to the noise parameters. In our workflow based on standard simulators (e.g., Qiskit Aer), such gradients are not natively available; obtaining them would require substantial reimplementation or numerical gradient estimation via finite differences, incurring $\mathcal{O}(d)$ additional objective evaluations per update. With $d=20$ parameters in our model, this imposes a substantial computational overhead given that the noisy simulation itself is the primary bottleneck. Additionally, the noise landscape is often non-convex and can be multimodal due to the complex interplay between coherent and incoherent error channels that challenges local search methods.
We thus employ Bayesian optimization to efficiently obtain estimates of $\boldsymbol{\theta}^*$. BO is a ML-based strategy well-suited for expensive-to-evaluate black-box objective functions~\cite{ArchettiCandelieri2019}. Unlike standard gradient-based local search, which determines step directions based on local derivatives, BO employs a probabilistic surrogate model to intelligently select parameter evaluations by maximizing an acquisition function that balances exploration of uncertain regions with exploitation of promising areas.

As a surrogate model, we use a tree-structured Parzen estimator \cite{bergstra2011algorithms}, which uses adaptive, non-parametric Parzen window density models to distinguish promising from poor regions of the search space based on past evaluations. Then samples new configurations from the most promising areas to enable efficient hyperparameter optimization in complex, hierarchical domains. We benchmark the BO against random search (RS) in which parameter points are randomly selected from the search space to provide unbiased model estimates. Both methods are implemented using Optuna \cite{optuna_2019}.

\subsection{Scalability analysis}

A key advantage of our approach is that we determine a fixed set of parameters within the noise model, irrespective of the circuit size or qubit count, leading to a circuit-size-independent complexity. This design choice decouples the computational complexity from the scale of the quantum hardware or the dimensions of the quantum system under study, enabling applicability to large-scale systems without prohibitive resource demands.

The resource requirements of our framework consist of experimental and computational costs. Experimentally, our data-efficient approach circumvents the prohibitive data overhead associated with full process tomography, which scales exponentially with qubit count due to the necessity of exhaustive state or process reconstructions. Instead, it relies solely on circuit-level outcome distributions obtained from a compact, predefined set of benchmark circuits executed on the target quantum hardware.
Computationally, the primary cost arises from the classical simulation of these benchmark circuits, required for each evaluation of the objective function during parameter determination. Notably, the cost per determination step remains constant, as it involves simulating the entire fixed set of training circuits to compute the discrepancy against the experimental dataset. Each iteration evaluates an objective function that quantifies this mismatch, here, via the Hellinger distance. The selection of the ML is pivotal for minimizing the number of such evaluations, given their computational intensity. We adopt BO, which excels in handling expensive black-box functions, accelerating convergence and enhancing the framework's scalability. Moreover, the entire characterization process is conducted as an offline, classical post-processing step, obviating any need for real-time interaction with the QPU.

\section{Experimental Results\label{sec:results}}

We benchmark our method using data collected from multiple IBM QPUs by executing quantum algorithms with varying qubit counts. The chosen systems ibmq\_kolkata, ibmq\_mumbai, and ibmq\_ehningen comprise 27 qubits with CNOT as the basis gate, physically implemented via echoed cross-resonance and single-qubit rotations. Notably, ibmq\_ehningen supports two distinct physical implementations of the CNOT gate \cite{ji2023optimizing}. In contrast, ibm\_kingston features 156 qubits and utilizes native CZ gates implemented via tunable couplers. Our noise model characterizes these basis gates, as all circuits must be decomposed into the basis gate set for execution.
The benchmarks include the QAOA and VQE, which are critical classes of variational algorithms designed for near-term hardware. We further include random circuits to evaluate the framework's suitability for unstructured quantum computations.
Moreover, we partition our benchmark circuits by size into a small-size training set (4--6 qubits) and a large-size validation set (7--9 qubits) for testing the model's ability to extrapolate from smaller systems to predict the behavior of larger, more complex systems.

\subsection{QAOA}

The QAOA constructs a parameterized quantum circuit consisting of alternating unitaries generated by a problem-dependent cost Hamiltonian $H_C$ and a mixing Hamiltonian $H_M$:
\begin{equation}
    U(\boldsymbol{\gamma}, \boldsymbol{\beta}) = \prod_{l=1}^L e^{-i\beta_l H_M} e^{-i\gamma_l H_C},
\end{equation}
where $\boldsymbol{\gamma}$ and $\boldsymbol{\beta}$ are variational parameters determined classically to minimize the expectation value of $H_C$. The mixing Hamiltonian is commonly chosen as $H_M=\sum_i X_i$, which promotes exploration of the Hilbert space. The cost Hamiltonian encodes the objective function of the optimization problem, typically taking the form of a diagonal operator in the computational basis
\begin{equation}
H_C = \sum_{i} c_i Z_i + \sum_{i<j} c_{ij} Z_i Z_j,
\end{equation}
where $Z_i$ are Pauli-$Z$ operators and the coefficients $c_i, c_{ij}$ reflect the problem structure. In this work, we employ QAOA applied to portfolio optimization problems, which involves fully connected two-qubit gates; further details can be found in \cite{ji2025algorithm,ji2023improving,ji2023optimizing,brandhofer2023benchmarking}.

For each QAOA circuit, we use the constrained optimization by linear approximation (COBYLA) \cite{powell1994direct} to choose angles $\boldsymbol{\gamma}$ and $\boldsymbol{\beta}$. Additionally, we use SWAP networks \cite{ji2025algorithm, montanez2025optimizing, weidenfeller2022scaling} on linear topologies to address the connectivity constraints of the QPU, which enables all compiled circuits to share the same fundamental gate structure and qubit connectivity pattern, regardless of the target device. This standardization allows us to control for variability in circuit design and compilation strategies.

We quantify circuit depth and gate counts of benchmark circuits executed on ibmq\_kolkata, ibmq\_mumbai, and ibmq\_ehningen to assess their complexity and noise sensitivity. As detailed in Table~\ref{tab:circuit_properties_kolkata}, the training set comprises circuits with 4--6 qubits and depths of $27.0 \pm 3.4$, while the prediction set exhibits 7--9 qubits and depths of $39.0\pm 3.4$. The scaling of CNOT gates, from $27.0 \pm 10.7$ in training to $78.0 \pm 18.4$ in prediction, reflects the increased entanglement complexity and noise accumulation in larger circuits. While the increase in circuit width (4--6 to 7--9 qubits) is moderate, this nearly $3\times$ increase in entangling gates drives the validation circuits into a significantly lower circuit fidelity regime (see Table \ref{tab:circuit_properties_qpus}), ensuring that the model is rigorously tested on its ability to generalize error dynamics into a more noise-affected operating regime distinct from the high circuit fidelity training regime.

\begin{table}[tb]
\caption{Properties of QAOA circuits executed on ibmq\_kolkata, ibmq\_mumbai, and ibmq\_ehningen. Metrics include circuit depth and gate counts (CNOT and single-qubit).}
\label{tab:circuit_properties_kolkata}
\begin{ruledtabular}
\begin{tabular}{lcc}
{Property} & {Training set} & {Prediction set} \\
\hline
Depth & $27.0 \pm 3.4$ & $39.0 \pm 3.4$ \\
CNOT gates & $27.0 \pm 10.7$ & $78.0 \pm 18.4$ \\
Single-qubit gates & $50.3 \pm 10.7$ & $92.3 \pm 13.2$ \\
\end{tabular}
\end{ruledtabular}
\end{table}

\subsubsection{Single-device characterization}

\begin{figure}[tb]
    \includegraphics[width=\columnwidth]{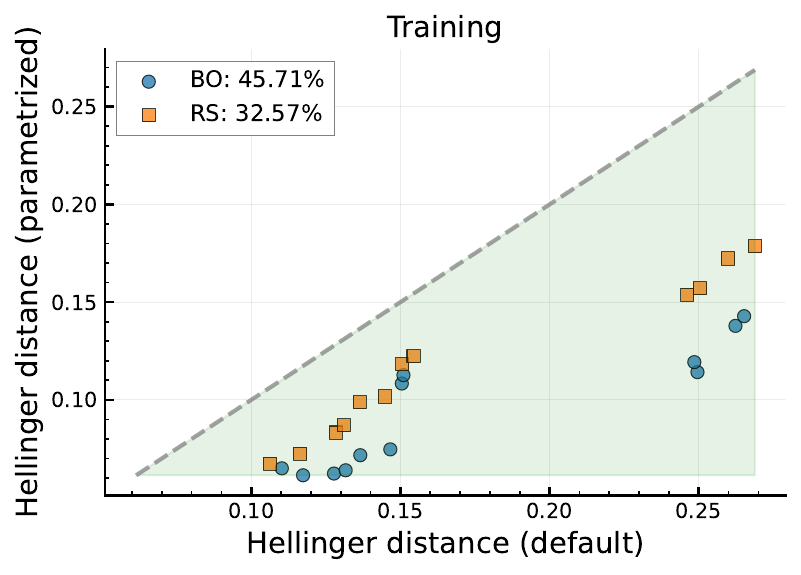}
    \put(-0.99\columnwidth, .70\columnwidth){\textbf{(a)}}\\
    \includegraphics[width=\columnwidth]{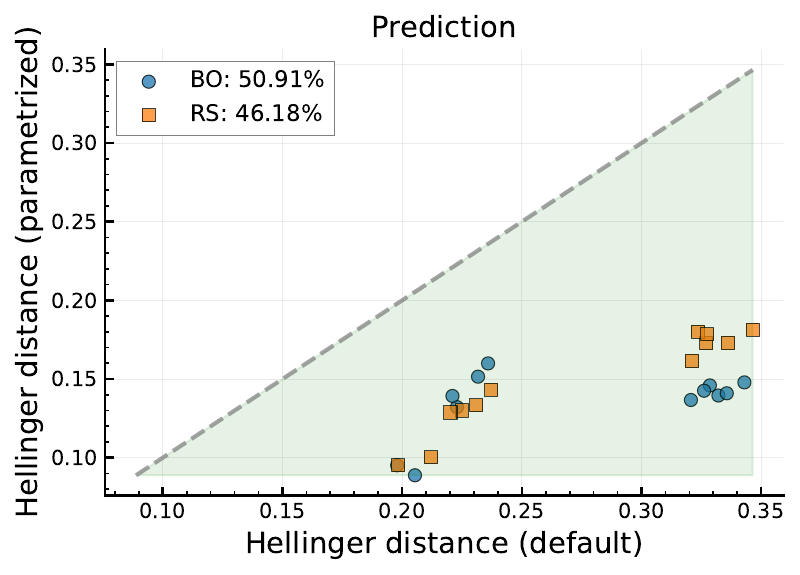}
    \put(-0.99\columnwidth, .70\columnwidth){\textbf{(b)}}\\
    \caption{Hellinger distances between outcomes from simulations using default device noise model ($x$-axis) and our parameterized noise models ($y$-axis) versus real executions on ibmq\_kolkata. Blue circles represent models optimized via BO, while orange squares denote those optimized through RS. All points below the dashed diagonal ($y=x$) indicate an improvement in model fidelity over the default model. The percentages shown in the legend represent the average reduction in Hellinger distance achieved by each optimization method relative to the default model. (a) Training phase using QAOA circuits on 4-, 5-, and 6-qubit instances, and (b) predictive performance on 7-, 8-, and 9-qubit circuits. Each circuit size was executed at four distinct times, yielding a total of 12 circuits for both training and prediction sets.}
    \label{fig:hd_kolkata}
\end{figure}

\begin{figure*}[tb]
    \includegraphics[width=\textwidth]{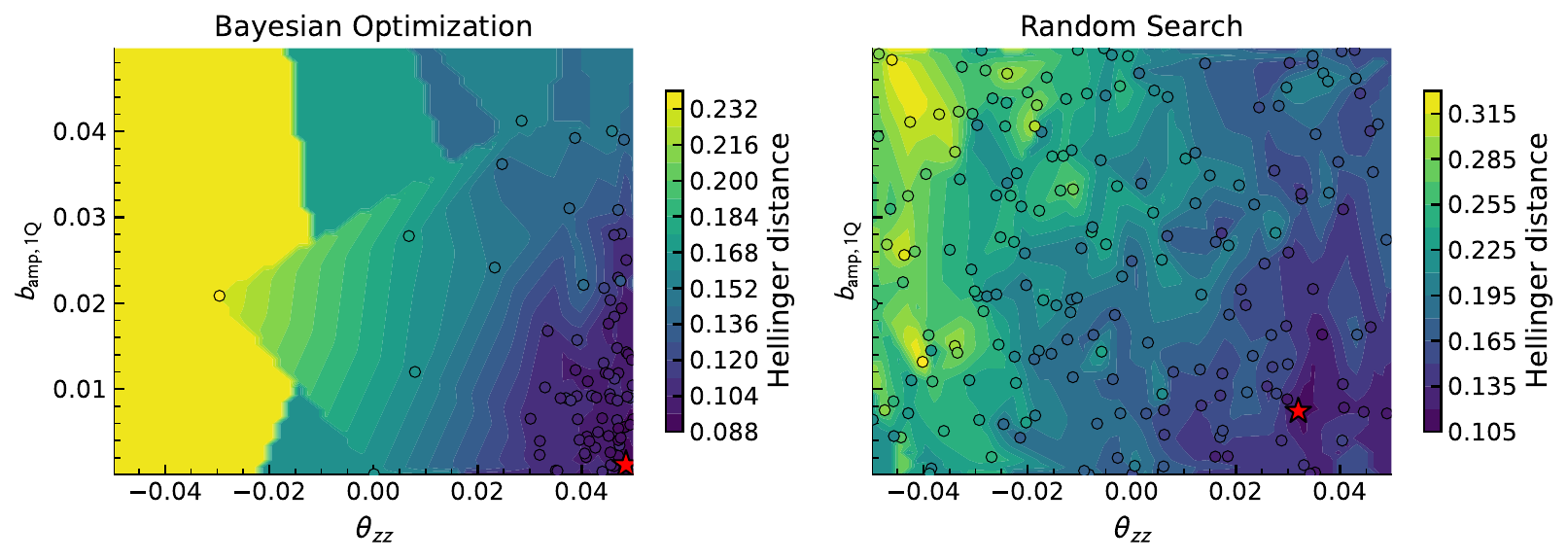}
    \put(-2.06\columnwidth, .70\columnwidth){\textbf{(a)}}
    \put(-\columnwidth, .70\columnwidth){\textbf{(b)}}
    \caption{Optimization landscape of the mean Hellinger distance, projected onto the two most significant noise parameters, as identified by Optuna, with marginalization over other noise model parameters. (a) Bayesian optimization and (b) random search. Each point represents a trial, and the red star marks the optimal configuration yielding the minimum value.}
    \label{fig:hd_kolkata_para_rela}
\end{figure*}

\begin{figure}[tb]
    \includegraphics[width=\columnwidth]{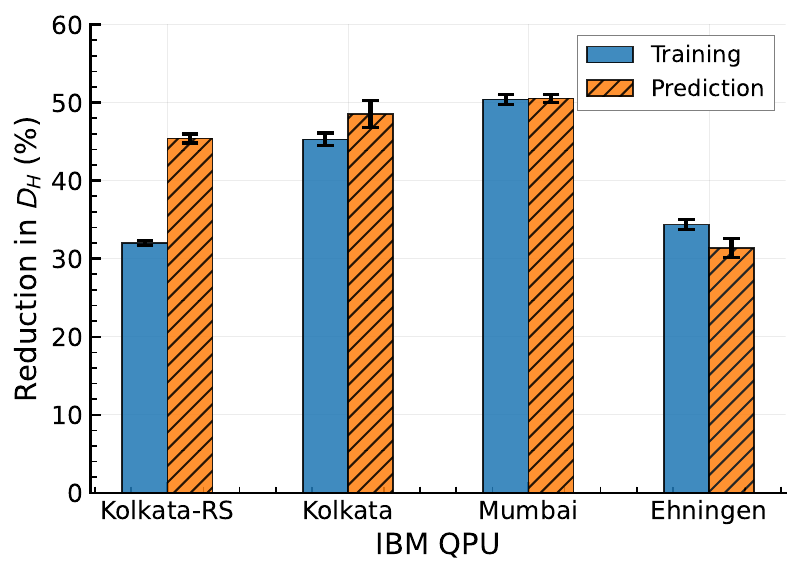}
    \put(-\columnwidth, .68\columnwidth){\textbf{(a)}}\\
    \includegraphics[width=\columnwidth]{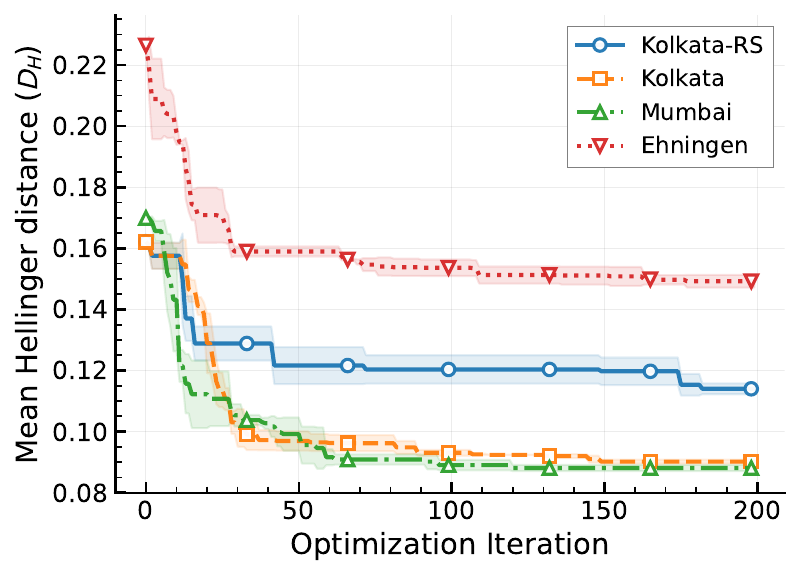}
    \put(-\columnwidth, .68\columnwidth){\textbf{(b)}}\\
    \caption{Comparison of BO and RS for training the parameterized noise model on ibmq\_kolkata (Kolkata-RS and Kolkata), alongside BO results on ibmq\_mumbai (Mumbai) and ibmq\_ehningen (Ehningen). (a) The average percentage reduction in Hellinger distance $D_H$ relative to the default noise model, for both training (blue, left) and prediction (orange, right) sets, evaluated on QAOA circuits with 4--6 qubits (training) and 7--9 qubits (prediction), each executed at four calibration instances. Error bars denote means and standard deviations derived from three independent repetitions across the 12 circuits per set. (b) The convergence of the optimization process corresponding to the training results presented in (a), which displays the best-achieved mean Hellinger distance of the training circuits over iterations.}
    \label{fig:tra_pre_qpus}
\end{figure}

We evaluate the method on ibmq\_kolkata.
On the training set, our parameterized noise model optimized with BO and RS achieves an average Hellinger distance of 0.095 and 0.118, respectively. This represents a significant reduction in predictive error of 45.71\% and 32.57\% compared to the default model's distance of 0.175, as shown in Fig.~\ref{fig:hd_kolkata}(a). Crucially, the model's performance generalizes robustly to the validation set (7–9 qubits) (Fig.~\ref{fig:hd_kolkata}(b)). For these larger circuits, our model achieves an average Hellinger distance of 0.135 (BO) and 0.148 (RS), corresponding to a reduction of 50.91\% and 46.18\% over the default model's distance of 0.275. These results also show that BO consistently outperforms RS in both the training and prediction processes, demonstrating its high optimization efficiency.

The contour plots in Fig.~\ref{fig:hd_kolkata_para_rela}(a) and~\ref{fig:hd_kolkata_para_rela}(b) compare the optimization landscapes explored by BO and RS, respectively, within a two-dimensional parameter space, which represent the most important parameters determined by Optuna. We observe that BO exhibits a structured and adaptive pattern, with sampled points increasingly concentrated in regions of minimal Hellinger distance, demonstrating that it efficiently identifies promising parameter regions. In contrast, RS yields a uniform and inefficient distribution of trials across the parameter space, resulting in a diffuse exploration that fails to converge effectively on the global minimum. This demonstrates the advantage of employing BO to achieve data-efficient parameterization.

\subsubsection{Cross-platform comparative analysis}

To assess the robustness of our data-efficient framework, we repeat the full training and prediction pipeline on ibmq\_kolkata three times, comparing BO to RS and plotting results with error bars in Fig.~\ref{fig:tra_pre_qpus}. As shown in Fig.~\ref{fig:tra_pre_qpus}(a), BO (Kolkata) outperforms RS (Kolkata-RS) in both training and prediction, exhibiting better performance. Moreover, Fig.~\ref{fig:tra_pre_qpus}(b) demonstrates that BO achieves a lower Hellinger distance than RS, confirming faster convergence to high-fidelity noise models.

In the following, we employ BO for determining the parameters of noise models and extend our analysis across two other QPUs (ibmq\_mumbai and ibmq\_ehningen) featuring varying noise characteristics to evaluate the generalizability and robustness of our approach.
For each device, we apply the same parameterized model architecture but train it independently on device-specific data. This process yields a unique model tailored to the specific error profile of each hardware instance.
As shown in Fig.~\ref{fig:tra_pre_qpus}, our approach consistently yields a substantial reduction in $D_H$ across all backends, and the optimization process converges to a low final $D_H$ value.
The framework's effectiveness is most pronounced on ibmq\_mumbai, which exhibits an average $D_H$ reduction of over 50\% for both training and prediction sets, with the reduction for individual prediction circuits peaking at 65\%. Strong performance is also observed on ibmq\_kolkata (45\% training,  49\% prediction) and ibmq\_ehningen (35\% training, 32\% prediction).
We summarize these achievements in Table~\ref{tab:circuit_properties_qpus}.
This high-fidelity performance across multiple systems, each with distinct error characteristics, confirms the robustness and broad applicability of our methodology for capturing hardware-specific noise profiles.

\begin{table*}[tb]
\caption{Experiment details of QAOA circuits executed on ibmq\_kolkata, ibmq\_mumbai, and ibmq\_ehningen. The circuit fidelity $F$ is defined in Eq.~\eqref{eq:circ_fid}. The physical qubits represent the active qubits selected for circuit execution within each benchmark set.}
\label{tab:circuit_properties_qpus}
\begin{ruledtabular}
\begin{tabular}{lcccccc}
& \multicolumn{2}{c}{ibmq\_kolkata} & \multicolumn{2}{c}{{ibmq\_mumbai}} & \multicolumn{2}{c}{{ibmq\_ehningen}} \\
\cline{2-3} \cline{4-5} \cline{6-7}
{Property} & Training & Prediction & Training & Prediction & Training & Prediction \\
\hline
Default $D_H$ & $0.175 \pm 0.000$ & $0.275 \pm 0.000$ & $0.179 \pm 0.001$ & $0.277 \pm 0.001$ & $0.230 \pm 0.000$ & $0.328 \pm 0.000$\\
Parameterized $D_H$ & $0.096 \pm 0.002$ & $0.141 \pm 0.004$ & $0.089 \pm 0.001$ & $0.137 \pm 0.001$ & $0.151 \pm 0.001$ & $0.225 \pm 0.004$ \\
$D_H$ reduction (\%) & $45.3 \pm 0.8$ & $48.5 \pm 1.7$ & $50.4 \pm 0.7$ & $50.5 \pm 0.5$ & $34.4 \pm 0.6$ & $31.3 \pm 1.2$\\
\hline
Circuit fidelity $F$ (\%) & $77.6 \pm 8.1$ & $47.9 \pm 8.3$ & $77.4 \pm 7.1$ & $51.5 \pm 6.7$ & $79.7 \pm 6.9$ & $53.2 \pm 8.0$ \\
Physical qubits & \makecell[c]{\{12, 15, 18, \\ 21, 23, 24\}} & \makecell[c]{\{4, 7, 10, 12, 15, \\ 18, 21, 23, 24\}} & \makecell[c]{\{14, 16, 19, 20, \\22, 24, 25\}} & \makecell[c]{\{12--16, 19, \\22, 24, 25\}} & \{15, 18--26\} & \{12--15, 18--26\} \\
Physical qubit count & 6 & 9 & 7 & 9 & 10 & 13\\
\end{tabular}
\end{ruledtabular}
\end{table*}

We use device calibration data to calculate the circuit fidelity $F$ to analyze the noise level of executing circuits on each device:
\begin{equation}
    F = \prod_i(1-\epsilon_i),
    \label{eq:circ_fid}
\end{equation}
where $\epsilon_i$ are gate and readout errors.
The increase in complexity from training set to prediction set corresponds to a significant circuit fidelity degradation from about $78\%$ to about $50\%$, as shown in Table~\ref{tab:circuit_properties_qpus}, underscoring the challenge of extrapolating noise models to larger circuit regimes where errors become more pronounced.
Notably, the physical qubit allocation, driven by device-specific transpilation strategies, reveals that prediction circuits span a broader device region compared to training circuits, potentially introducing greater variability from spatial heterogeneity in device error rates.
Moreover, the subgraph of active qubits is not fixed from circuit to circuit, further increasing the diversity of noise conditions encountered during evaluation.
We observe that {ibmq\_ehningen}, compared to other QPUs, maps circuits across a significantly wider physical qubit topology across four different benchmark executions.
The increased spatial complexity is probably responsible for the more modest performance gains. A potential reason is that our amount of training data is insufficient to fully characterize the more complex error landscape of an expanded qubit ensemble.
Nevertheless, the systematic increase in circuit complexity from training to prediction provides a meaningful test of our noise model's ability to extrapolate beyond its training regime.
This key extrapolation is enabled because our framework avoids qubit-specific error rate mappings. Instead, it constructs a generative model of the device's noise, optimizing global hyperparameters $\boldsymbol{\theta}$ that parameterize statistical distributions for gate errors from which individual qubit and gate parameters are drawn.

In our experiments, we maintain circuit fidelities $F \gtrsim 50\%$ in both the training phase (4–6 qubits) and the larger-scale prediction phase (7–9 qubits), so their output distributions remain far from uniform and retain structure that is sensitive to coherent and correlated error mechanisms. By contrast, circuits with substantially larger width (e.g., 20–30 qubits) at comparable depth on the same devices are expected to operate close to the noise floor, yielding nearly uniform outcome distributions that obscure detailed differences between noise models and strongly reduce the discriminative power of metrics such as the Hellinger distance. Our conclusions should therefore be understood as applying to this intermediate-fidelity regime; extending the analysis to significantly larger mapped circuits will require dedicated experiments and is left for future work.

\begin{figure}[tb]
    \includegraphics[width=\columnwidth]{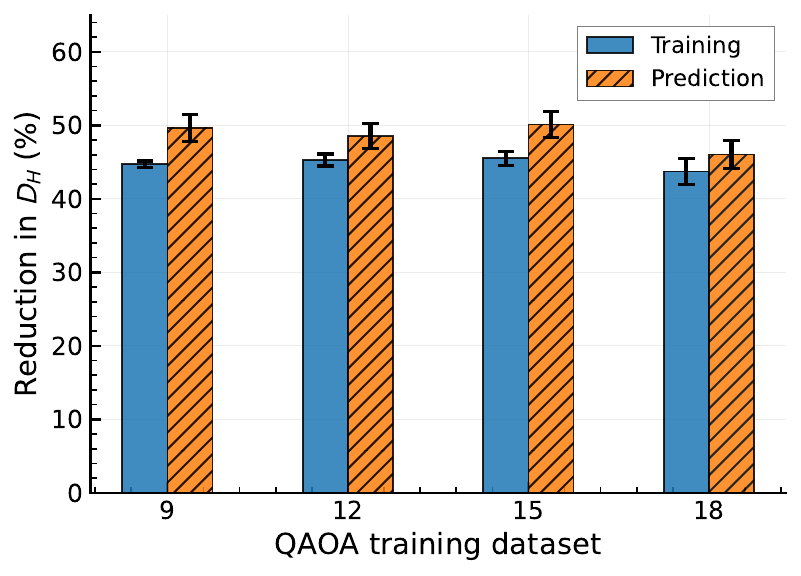}
    \caption{Impact of different training dataset sizes on the performance of BO-optimized noise model on ibmq\_kolkata. Each training set is denoted by the number of QAOA circuits. The experimental setting is the same as in Fig.~\ref{fig:tra_pre_qpus}.}
    \label{fig:tra_pre_data_points}
\end{figure}

\subsubsection{Training efficiency study}

This section examines the interplay between model performance and training dataset size to identify the minimal data volume needed for target accuracy with reduced resource demands.
To assess our framework's data requirements, we train and validate our noise model using four distinct QAOA datasets of varying sizes (9, 12, 15, and 18 circuits). The results reveal strong robustness and data efficiency.
Across training-set sizes from 9 to 18 circuits, circuit properties remain stable. The preserved complexity distribution across all dataset sizes, combined with consistent fidelity estimates, suggests that our smallest 9-circuit training set sufficiently constrains the model for this benchmark; more details can be found in Table~\ref{tab:circuit_properties_kolkata_scaling} in Appendix~\ref{app:exp_detail}. This structural invariance explains the robust performance shown in Fig.~\ref{fig:tra_pre_data_points}, where the framework achieves a consistently similar relative reduction in the mean $D_H$ of approximately 45\% to 50\% across all tested datasets, despite a $2\times$ difference in training data volume.
Moreover, the optimization convergence profiles for 12, 15, and 18 circuit datasets are nearly indistinguishable, reaching the same optimal $D_H$ value at a similar rate, highlighting stability (see Fig.~\ref{fig:tra_pre_data_points_opt} in Appendix~\ref{app:exp_detail}). 

These results demonstrate that the framework is data-efficient and provides empirical evidence of performance saturation with respect to training-set size, with no indication of overfitting in the tested regime. Within the tested range (9--18 training circuits), performance stabilizes: increasing the training set from 9 to 18 circuits yields no appreciable change in the optimal $D_H$ and produces nearly indistinguishable convergence behavior (Fig.~\ref{fig:tra_pre_data_points_opt}). This is consistent with adequate constraint diversity: circuits with varying depths and two-qubit-gate densities probe distinct error channels, providing complementary constraints on the shared 20-parameter model, while physics-informed functional forms restrict effective capacity by confining $\boldsymbol{\theta}$ to physically realistic mechanisms. Out-of-distribution validation, where improvements match or in several cases exceed training across devices, further confirms robust generalization.

\subsection{VQE\label{subsec:vqe}}

To verify the efficacy of the method beyond QAOA, we perform a similar optimization for VQE circuits on ibm\_kingston, a 156-qubit Heron R2 
processor featuring IBM's heavy-hex topology. We consider the task of minimizing the ground state energy of the cluster Hamiltonian
\begin{equation}
    H_n = \alpha_0Z_0 + \alpha_nZ_n + \sum_{j=0}^{n-3}\beta_jZ_jX_{j+1}Z_{j+2}\,,
\end{equation}
where $\alpha_i$ and $\beta_i$ are real coefficients.
In the following, we choose $\alpha_k=1/n$ ($\beta_k=1/n$) for all $k$ for $n=4,\dots,9$. Using a simple, hardware-efficient ansatz corresponding to the \texttt{EfficientSU2} circuit in Qiskit, we optimize the parameters using the sequential least squares quadratic programming (SLSQP) \cite{kraft1988software} in a noiseless simulation and verify that the optimized circuits converge to the true ground state energy. We then execute the optimized circuits on the 156-qubit Heron R2 QPU {ibm\_kingston}. For the execution, we group the circuits for each $n$ into two qubit-wise commuting groups and sample from each group with $30.000$ shots. This results in a total of $12$ circuits which we partition into six circuits ($n=4,5,6$) for training and six circuits for validation ($n=7,8,9$).

As in the QAOA study, we apply the same parameterized noise model architecture to the VQE experiments on ibm\_kingston; however, the parameters $\boldsymbol{\theta}$ are learned solely from the VQE training data specific to ibm\_kingston.
Performing three independent optimizations using $200$ iterations each, we obtain optimized parameters which result in an improvement of $43.99\%\pm0.14\%$ in model fidelity for the training set and $18.58\%\pm0.90\%$ for the validation set (up to 27\% for a circuit prediction, see Fig.~\ref{fig:app_vqe_rc}(a) in Appendix~\ref{app:exp_detail}). While the parameterized model aligns more closely with experiment, especially on the training circuits, the improvement on the validation set is less pronounced than in the QAOA case. This likely reflects differences between the devices from which the data were sampled. As detailed in Table~\ref{tab:circuit_properties_kingston}, VQE circuits contain significantly fewer entangling gates (CZs) than QAOA circuits (CNOTs). Moreover, the ibm\_kingston processor itself exhibits lower error rates. These factors contribute to inherently higher baseline circuit fidelities, resulting in substantially smaller Hellinger distances between the baseline model and experiment. On the validation set, the default model’s Hellinger distance is less than half of that observed for the comparable model on the older generation devices used for QAOA. Consequently, the margins of improvement over the baseline are smaller. Nevertheless, these results demonstrate that our framework is robust across diverse hardware architectures, accommodating distinct qubit counts and native gate sets, establishing a clear path forward for modeling noise in more complex VQE circuits.

\begin{table}[tb]
\caption{Properties of VQE circuits executed on ibm\_kingston.}
\label{tab:circuit_properties_kingston}
\begin{ruledtabular}
\begin{tabular}{lcc}
{Property} & {Training set} & {Prediction set} \\
\hline
Default $D_H$ & $0.075 \pm 0.000$ & $0.124 \pm 0.000$ \\
Parameterized $D_H$ & $0.042 \pm 0.000$ & $0.101 \pm 0.001$ \\
$D_H$ reduction (\%) & $43.99\%\pm0.14\%$ & $18.58\%\pm0.90\%$\\
\hline
Depth & $41.2 \pm 3.2$ & $53.7 \pm 3.4$ \\
CZ gates & $12.0 \pm 2.7$ & $21.0 \pm 2.7$ \\
Single-qubit gates & $107.2 \pm 21.9$ & $184.7 \pm 19.3$ \\
Circuit fidelity $F$ (\%) & $91.0 \pm 1.5$ & $83.8 \pm 3.9$ \\
Physical qubits & \makecell[c]{\{0--5\}} & \makecell[c]{\{0--8\}} \\
Physical qubit count & 6 & 9\\
\end{tabular}
\end{ruledtabular}
\end{table}

\subsection{Random circuits\label{subsec:rc}}

We further benchmark the method through random circuits on ibmq\_kolkata.
Random quantum circuits were generated using Qiskit's \texttt{random\_circuit} function with systematically varied parameters: qubit count $n \in \{4, ..., 10\}$ and depth $d = n + 3$. The function constructs circuits by randomly sampling single-qubit gates from SU(2) and placing two-qubit gates between random qubit pairs within the depth constraint, with terminal measurements on all qubits. 
The results show a substantial mean $D_H$ improvement of 43.60\% in model fidelity for training and 45.90\% for prediction (see Fig.~\ref{fig:app_vqe_rc}(b) in Appendix~\ref{app:exp_detail}). We perform three independent optimization processes and achieve a mean Hellinger distance reduction of $42.9\% \pm 1.5\%$ for the training set and $46.8\% \pm 2.1\%$ for the prediction set, with a maximum reduction of 63\% observed for a circuit in the prediction set.
This demonstrates that our method's high performance is a statistically significant result, making it a reliable tool for practical application-aware hardware characterization.

\begin{table}[tb]
\caption{Properties of random circuits executed on ibmq\_kolkata.}
\label{tab:circuit_properties_rc}
\begin{ruledtabular}
\begin{tabular}{lcc}
{Property} & {Training set} & {Prediction set} \\
\hline
Default $D_H$ & $0.162 \pm 0.000$ & $0.250 \pm 0.000$ \\
Parameterized $D_H$ & $0.093 \pm 0.002$ & $0.133 \pm 0.005$ \\
$D_H$ reduction (\%) & $42.9 \pm 1.3$ & $46.8 \pm 1.9$\\
\hline
Depth & $87.2 \pm 18.9$ & $182.6 \pm 23.8$ \\
CNOT gates & $38.3 \pm 17.8$ & $136.6 \pm 20.4$ \\
Single-qubit gates & $106.8 \pm 32.0$ & $271.7 \pm 34.3$ \\
Circuit fidelity $F$ (\%) & $64.0 \pm 14.0$ & $24.8 \pm 7.8$ \\
Physical qubits & \makecell[c]{\{0--7, 10, 12, 15, \\ 17, 18, 21, 23, 24\}} & \makecell[c]{\{0, 1, 4, 6, 7, 10, \\ 12--19, 21, 23, 24\}} \\
Physical qubit count & 16 & 17 \\
\end{tabular}
\end{ruledtabular}
\end{table}

Table~\ref{tab:circuit_properties_rc} highlights the markedly distinct complexity profile of the random circuits. Compared to QAOA benchmarks, these random circuits exhibit approximately 3$\times$ greater depth and employ a broader, more fragmented qubit allocation across 17 physical qubits on the device. This escalation in depth and spatial distribution leads to substantially degraded fidelities, $64.0\%\pm14.0\%$ for the training circuits, and $24.8\%\pm7.8$ for the prediction sets. The 3.5$\times$ scaling in CNOT gates from training to prediction surpasses the 2.9$\times$ factor observed in QAOA, underscoring the amplified accumulation of noise in deeper, less structured entangling operations. Remarkably, under these stringent conditions with fidelities roughly 25\%, our approach achieves a performance improvement of around 44\%, affirming its exceptional robustness across varied circuit topologies and noise regimes inherent to superconducting quantum hardware.

\subsection{Application-aware noise characterization\label{subsec:appli_aware_noise}}

The results show that the proposed framework achieves high predictive accuracy across three qualitatively different algorithmic families (QAOA, VQE, and random circuits). Crucially, rather than seeking a single universal parameter vector $\boldsymbol{\theta}$ that simultaneously describes all algorithm classes on a target device, we adopt an application-aware philosophy: we estimate independent parameter sets for each algorithm-hardware context while retaining a fixed model architecture. This design is physically motivated by the context-dependent nature of noise in superconducting processors. Different algorithms can emphasize different effective error mechanisms through their circuit structure (e.g., depth, routing, and parallelism). For instance, QAOA applied to portfolio-optimization problems can require extensive SWAP-network routing to achieve all-to-all connectivity, whereas VQE with hardware-efficient ansätze typically conforms to the native coupling map, resulting in significantly shallower depths and fewer entangling gates. Since our parameterization captures duration-aware dephasing and correlated errors, we do not expect a parameter set calibrated on high-depth QAOA circuits to accurately predict low-depth VQE fidelity, or vice versa. Generality is thus achieved at the framework level: the same workflow applies across devices and algorithms, while the effective parameters adapt to the specific error dynamics of each context.

While cross-algorithm parameter transfer is not a prerequisite for deployment, quantifying it would elucidate the fundamental trade-off between intrinsic device physics and application-specific error manifestations. Investigating the predictive validity of a model trained on one algorithm family (e.g., QAOA) when applied to another (e.g., VQE) helps distinguish between static hardware properties and the effective error dynamics induced by specific circuit designs.
Future work utilizing transfer learning could quantify the adaptation cost between these contexts, ultimately determining whether the fidelity benefits of application-aware calibration justify the overhead of independent training or if approximate universality offers a viable compromise for large-scale systems.

\section{Discussion and Conclusion\label{sec:conclusion}}

This work develops a novel parameterized noise model with circuit-size-independent complexity. Leveraging ML-driven optimization, we repurpose data from a few routine circuit executions to construct data-efficient and prediction accurate noise models for superconducting quantum processors.
Crucially, our framework prioritizes application-aware characterization: we estimate effective noise parameters independently for each algorithm family on the target hardware rather than seeking a universal parameter set across all algorithm families. This design captures the distinct interaction between the algorithm and hardware, achieving high predictive accuracy with no additional dedicated calibration circuits or QPU jobs beyond the workload executions themselves.
By systematically searching complex, multi-parameter error spaces, our approach consistently produces models that are significantly more predictive than default ones obtained from hardware vendors.
We demonstrate a reduction of up to 65\% in the Hellinger distance, a metric that quantifies the divergence between predicted and experimental measurement-outcome (bitstring) probability distributions, confirming that our optimized models produce output distributions that more accurately represent the behavior of algorithms executed on real quantum hardware.
This significant improvement in predictive accuracy is essential for a reliable simulation of quantum circuits on noisy quantum devices, providing a foundation for more accurate noise-aware quantum circuit compilation by enabling informed optimization of qubit selection, SWAP gate insertion, and other compilation passes.

A key advantage of our approach is its efficiency and practicality. It repurposes existing benchmarking or application-specific execution data, avoiding the need for extensive characterization protocols such as quantum process tomography.
This reusability reduces experimental overhead, enabling practitioners to leverage prior executions to improve algorithm testing and error mitigation protocols while iteratively refining noise models as additional data become available.
Comprehensive benchmarking across multiple IBM quantum devices and quantum algorithms validates the framework’s adaptability to different types of superconducting hardware, highlighting its generality and potential for broad adoption in applications. By bridging the gap between theoretical models and experimental reality, our data-efficient approach provides a critical enabling technology for developing the next generation of effective error mitigation techniques and noise-aware compilers, accelerating the realization of practical quantum computing applications.

Future work should address scaling challenges through hierarchical noise modeling. As system size grows, achieving a target accuracy may require more expressive parameterizations, increasing both the ML optimization cost and the amount of circuit-execution data. One promising direction is to partition the device coupling graph into regional subgraphs and assign region-specific parameter sets, capturing spatial inhomogeneity in larger systems while avoiding per-component overparameterization.
Moreover, as differentiable or surrogate-based simulators mature, it will be valuable to investigate gradient-based and hybrid strategies, such as BO for initialization followed by gradient refinement, to further improve scalability in higher-dimensional parameter spaces.
Additionally, investigating transfer learning across algorithmic families could quantify the adaptation cost between algorithmic contexts, helping determine if approximate universality is a viable compromise for large-scale systems. Moreover, incorporating leakage errors, where qubits excite outside of the computational subspace, represents a critical extension. This requires higher-dimensional simulations and often specialized characterization protocols, substantially increasing the computational burden of the ML loop. Although beyond the scope of this initial study, incorporating leakage is essential for further fidelity gains.

Our pairwise readout correlation model captures nearest-neighbor effects but neglects long-range correlations; small cluster-wise confusion matrices tied to resonators or readout lines could provide systematic improvements. Benchmarking dense algorithms like QAOA on utility-scale processors would complement our VQE results  on 156-qubit hardware, revealing how the framework interacts with 
large-scale qubit mapping strategies.
Additionally, extending to alternative qubit modalities represents essential future work. While our experiments focus on superconducting devices, the methodology is architecture-agnostic. Adaptation requires replacing the physics-informed ansatz $\mathcal{N}(\boldsymbol{\theta})$ with platform-specific mechanisms. For trapped-ion processors \cite{bruzewicz2019trapped,moses2023arace}, for instance, a tailored ansatz would capture dominant error sources in Mølmer–Sørensen gates, such as motional mode heating and laser intensity fluctuations.
The ultimate goal is a full integration into quantum compiler toolchains, enabling adaptive, algorithm-specific compilation that can dynamically tailor noise models to both the hardware's current state and circuit structure. This would optimize execution by capturing the unique interplay between algorithms and device-specific error channels.

\section*{Acknowledgment}

The paper was written as part of the project ``Quantum-based Energy Grids (QuGrids)", which is receiving funding from the programme ``Profilbildung 2022", an initiative of the Ministry of Culture and Science of the State of North Rhine-Westphalia.
The contribution of M. R. and D. A. K. has been supported by the Fraunhofer Heilbronn Research and Innovation Centers HNFIZ.
Parts of the data used for this study have been generated in the course of project QORA funded by by Ministerium f\"{u}r Wirtschaft, Arbeit und Tourismus Baden-W\"{u}rttemberg in the frame of the Competence Center Quantum Computing Baden-Württemberg.
We acknowledge the use of IBM Quantum services for this work and to advanced services provided by the IBM Quantum Researchers Program. The views expressed are those of the authors, and do not reflect the official policy or position of IBM or the IBM Quantum team.

\appendix

\section{Experimental details\label{app:exp_detail}}

\begin{figure}[tb]
    \includegraphics[width=\columnwidth]{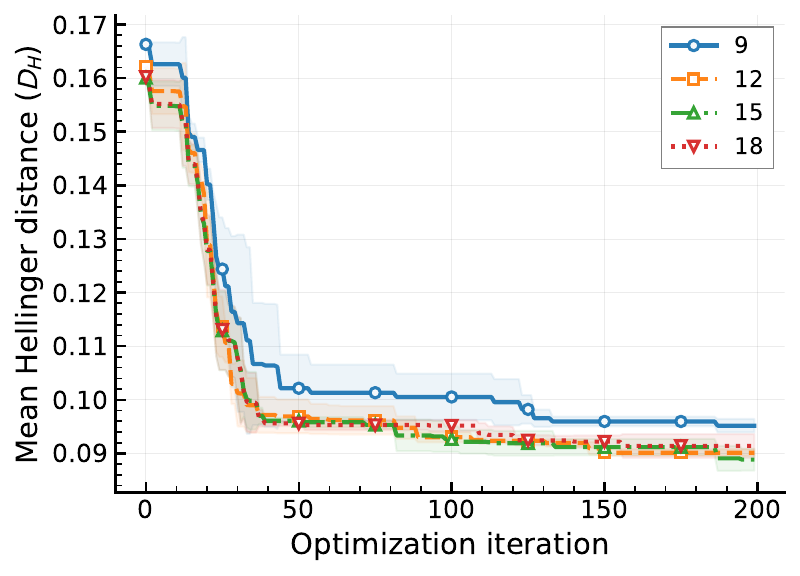}
    \caption{Convergence of the optimization process across different circuit sizes on ibmq\_kolkata in Fig.~\ref{fig:tra_pre_data_points}.}
    \label{fig:tra_pre_data_points_opt}
\end{figure}

We present the experimental details.
Figure~\ref{fig:tra_pre_data_points_opt} shows the optimization convergence across different QAOA circuit sizes on ibmq\_kolkata, corresponding to Fig.~\ref{fig:tra_pre_data_points}.
Table~\ref{tab:circuit_properties_kolkata_scaling} quantifies the structural consistency of our benchmark circuits across different training set sizes.
The physical qubits used are identical across different circuit sizes: \{12, 15, 18, 21, 23, 24\} for the training set and \{4, 7, 10, 12, 15, 18, 21, 23, 24\} for the prediction set.

\begin{table*}[tb]
\caption{\label{tab:circuit_properties_kolkata_scaling}Properties of benchmark circuit sets on {ibmq\_kolkata}, analyzed for different numbers of QAOA circuits in Fig.~\ref{fig:tra_pre_data_points}.}
\begin{ruledtabular}
\begin{tabular}{lcccccccc}
 & \multicolumn{4}{c}{Training set} & \multicolumn{4}{c}{Prediction set} \\
\cmidrule(lr){2-5} \cmidrule(lr){6-9}
Property & 9 circuits & 12 circuits & 15 circuits & 18 circuits & 9 circuits & 12 circuits & 15 circuits & 18 circuits \\
\hline
Depth & $27.0 \pm 3.5$ & $27.0 \pm 3.4$ & $27.0 \pm 3.4$ & $27.0 \pm 3.4$ & $39.0 \pm 3.5$ & $39.0 \pm 3.4$ & $39.0 \pm 3.4$ & $39.0 \pm 3.4$ \\
CNOT gates & $27.0 \pm 10.9$ & $27.0 \pm 10.7$ & $27.0 \pm 10.6$ & $27.0 \pm 10.5$ & $78.0 \pm 18.6$ & $78.0 \pm 18.4$ & $78.0 \pm 18.2$ & $78.0 \pm 18.1$ \\
Single-qubit gates & $50.3 \pm 10.8$ & $50.3 \pm 10.7$ & $50.3 \pm 10.6$ & $50.3 \pm 10.5$ & $92.3 \pm 13.4$ & $92.3 \pm 13.2$ & $92.3 \pm 13.1$ & $92.3 \pm 13.0$ \\
Circuit fidelity $F$ (\%) & $77.3 \pm 8.1$ & $77.6 \pm 8.1$ & $77.7 \pm 8.1$ & $77.8 \pm 8.1$ & $48.3 \pm 8.2$ & $47.9 \pm 8.3$ & $47.7 \pm 8.3$ & $47.5 \pm 8.3$ \\
\end{tabular}
\end{ruledtabular}
\end{table*}

\begin{figure*}[tb]
    \includegraphics[width=.48\textwidth]{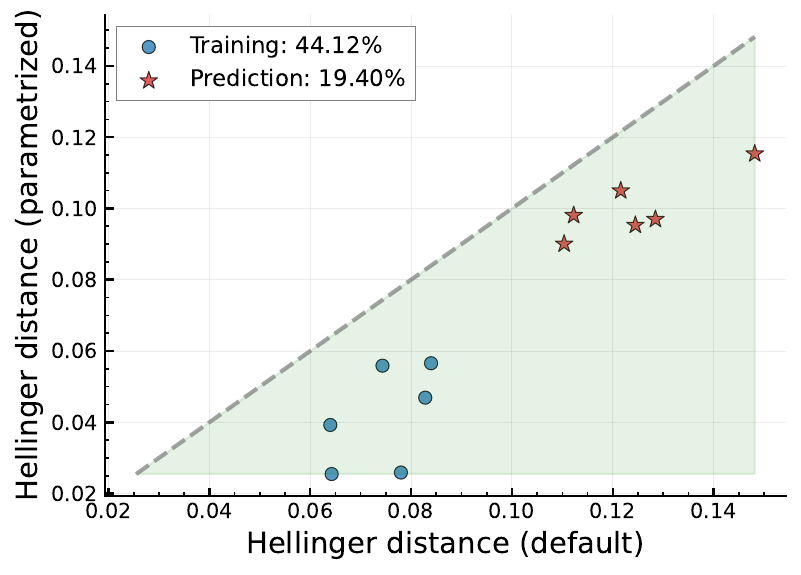}\hfill
    \includegraphics[width=.48\textwidth]{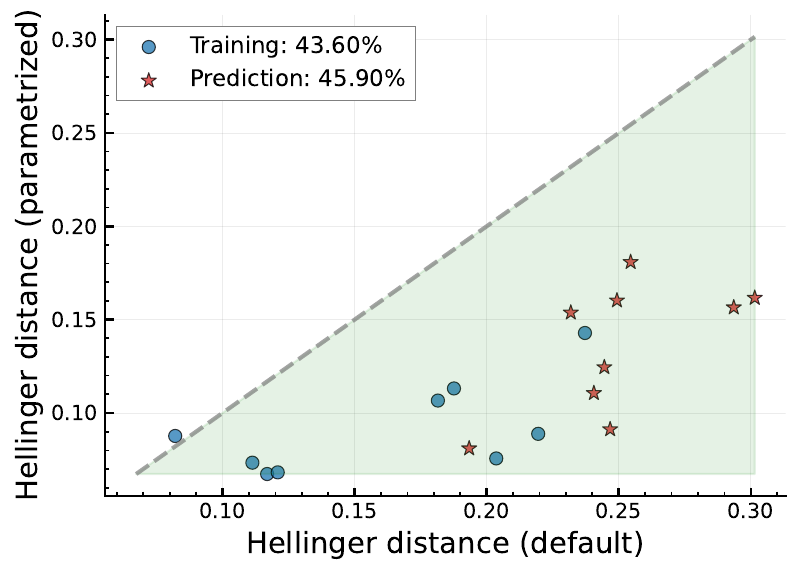}
    \put(-1.15\columnwidth, .70\columnwidth){\textbf{(a)}}
    \put(-0.105\columnwidth, .70\columnwidth){\textbf{(b)}}
    \caption{Performance comparison of noise models using (a) VQE on ibm\_kingston and (b) random circuits on {ibmq\_kolkata}.}
    \label{fig:app_vqe_rc}
\end{figure*}

Figure~\ref{fig:app_vqe_rc}(a) compares our parameterized model against the default model using VQE obtained on ibm\_kingston in Sec.~\ref{subsec:vqe}, while Fig.~\ref {fig:app_vqe_rc}(b) presents results using random circuits on ibmq\_kolkata in Sec.~\ref{subsec:rc}.

\bibliography{refs}

\end{document}